\documentclass[preprint,12pt]{elsarticle}

\usepackage{amssymb}
\usepackage{amsmath}

\usepackage{threeparttable}
\usepackage[english]{babel}
\usepackage[T1]{fontenc}
\usepackage[utf8]{inputenc}
\usepackage{booktabs} 
\usepackage[table]{xcolor}      

\usepackage{acronym}

\usepackage{hyperref}
\hypersetup{
  colorlinks   = true, 
  urlcolor     = blue, 
  linkcolor    = black, 
  citecolor   = blue 
}

\usepackage{datatool}

\journal{Elsevier}

\begin{document}

\begin{frontmatter}



\title{\textbf{Challenges and Opportunities for Second-life Batteries: A Review of Key Technologies and Economy}}


\author{Xubo Gu$^{a}$, Hanyu Bai$^{a}$, Xiaofan Cui$^{b}$, Juner Zhu$^{c}$, Weichao Zhuang$^{d}$, Zhaojian Li$^{e}$, Xiaosong Hu$^{f}$, and Ziyou Song$^{a*}$}

\cortext[mycorrespondingauthor]{Corresponding Authors:  ziyou@nus.edu.sg (Z. Song)}

\address{$^a$Department of Mechanical Engineering, National University of Singapore, Singapore, 117575, Singapore}
\address{$^b$Department of Electrical Engineering and Computer Science, University of Michigan, Ann Arbor, MI, 48109, USA}
\address{$^c$Department of Mechanical and Industrial Engineering, Northeastern University, Boston, MA 02115, USA}
\address{$^d$School of Mechanical Engineering, Southeast University, Nanjing, 211189, China}
\address{$^e$Department of Mechanical Engineering, Michigan State University, East Lansing, MI, 48824, USA}
\address{$^f$Department of Mechanical and Vehicle Engineering, Chongqing University, Chongqing, 400044, China}


\begin{abstract}\small
Due to the increasing volume of Electric Vehicles in automotive markets and the limited lifetime of onboard lithium-ion batteries (LIBs), the large-scale retirement of LIBs is imminent. The battery packs retired from Electric Vehicles still own 70\%-80\% of the initial capacity, thus having the potential to be utilized in scenarios with lower energy and power requirements to maximize the value of LIBs.  
However, spent batteries are commonly less reliable than fresh batteries due to their degraded performance, thereby necessitating a comprehensive assessment from safety and economic perspectives before further utilization.
To this end, this paper reviews the key technological and economic aspects of second-life batteries (SLBs). 
Firstly, we introduce various degradation models for first-life batteries and identify an opportunity to combine physics-based theories with data-driven methods to establish explainable models with physical laws that can be generalized. However, degradation models specifically tailored to SLBs are currently absent. Therefore, we analyze the applicability of existing battery degradation models developed for first-life batteries in SLB applications.
Secondly, we investigate fast screening and regrouping techniques and discuss the regrouping standards for the first time to guide the classification procedure and enhance the performance and safety of SLBs.
Thirdly, we scrutinize the economic analysis of SLBs and summarize the potentially profitable applications.  
Finally, we comprehensively examine and compare power electronics technologies that can substantially improve the performance of SLBs, including high-efficiency energy transformation technologies, active equalization technologies, and technologies to improve reliability and safety.
\end{abstract}



\end{frontmatter}


\section*{Acronyms}
\begin{acronym}[AWGN]
\acro{AC}{Alternating Current}
\acro{BESS}{Battery Energy Storage System}
\acro{COE}{Cost of Energy}
\acro{DDM}{Data-driven Model}
\acro{DSPV}{Distributed Solar Photovaltaic}
\acro{DC}{Direct Current}
\acro{EV}{Electric Vehicle}
\acro{EOL}{End-of-life}
\acro{ESS}{Energy Storage System}
\acro{FLB}{First-life Battery}
\acro{FPP}{Full Power Processing}
\acro{LIB}{Lithium-ion Battery}
\acro{LLI}{Loss of Lithium Inventory}
\acro{acroym}[LAM\textsubscript{NE}]{Loss of Anode Active Material}
\acro{acroym}[LAM\textsubscript{PE}]{Loss of Cathode Active Material}
\acro{LCOE}{Levelized Cost of Electricity}
\acro{LCOS}{Levelized Cost of Storage}
\acro{MMC}{Multilevel Modular Converters}
\acro{NE}{Negative Electrode}
\acro{NPV}{Net Present Value}
\acro{PBM}{Physics-based Model}
\acro{PINN}{Physics-informed Neural Network}
\acro{PE}{Positive Electrode}
\acro{PDE}{Partial Differential Equation}
\acro{PPP}{Partial Power Processing}
\acro{PV}{Photovoltaic}
\acro{SLB}{Second-life Battery}
\acro{SEI}{Solid Electrolyte Interphase}
\acro{SOH}{State of Health}
\end{acronym}

\section{Introduction}
Electric Vehicles (EVs) have become increasingly popular in recent years. In 2022, global sales of EVs reached 10.5 million units, accounting for 13\% of total light vehicle sales. This represents a 55\% increase from 2021, indicating a strong growth trend \cite{roland_ev-volumes_2023}. 
Most commercial EVs adopt lithium-ion batteries (LIBs) because of their excellent properties, such as high energy density and high power density. Typically, the lifespan of the LIB pack in an EV is around 8-10 years, after which the battery is retired when its remaining capacity decreases to 70\%-80\% of its initial value. Due to the high volume of EVs being in service and the limited lifespan of LIBs, a significant volume of retired batteries is expected in the near future.
To reduce the cost of EVs and mitigate their environmental impacts, the retired LIBs should be reused and ultimately recycled. These retired batteries can still retain 70\%-80\% of their original capacity and can be utilized in scenarios with lower energy and power requirements, such as energy storage stations or communication base stations \cite{Lai2021TurningBatteries}. In this way, the value of LIBs can be maximized in their second-life applications. Meanwhile, the upfront cost of EVs will be reduced since car owners can recover some of their value from selling their retired batteries. The metals and electrolytes in LIBs would cause heavy metal and chemical pollution to the environment if they were directly discarded \cite{Harper2019RecyclingVehicles}. Hence, it is important to recycle LIBs after their echelon utilization to extract valuable materials for further battery production. The development of an effective echelon utilization and recycling system is crucial to support the sustainable growth of the EV industry and has broad societal significance worldwide. 

However, the effective utilization of second-life batteries (SLBs) is a multifaceted problem.  
Firstly, the determination of SLB's internal status is complicated. The status of SLBs consists of the internal state (e.g., SEI layer and lithium plating), the external characteristics (e.g., capacity and power capability), and the remaining life expectancy. Note that measuring and characterizing the internal changes of SLBs is generally not straightforward, as different SLBs from various EVs have experienced different operational conditions, resulting in varying internal statuses in anodes, electrolytes, cathodes, and current collectors, thereby making it difficult to establish common testing protocols to assess the status of SLBs.  For example, the used battery cells with the same capacity may have significantly different remaining life due to various dominant degradation mechanisms inside. Nondestructive detection methods, such as in-situ neutron diffraction, have been used in academia \cite{wandt_quantitative_2018}. However, their widespread application in large-scale retired LIBs can be costly. 
Secondly, reusing a large number of retired batteries will lead to enormous demand for the disassembly and sorting of heterogeneous battery packs/modules/cells, which can be labor-intensive and expensive. Moreover, sorting batteries becomes challenging due to the loss of historical data, which necessitates extensive new testing, increasing the cost of SLBs. 
Thirdly, from an economic point of view, the profitability of SLBs in different applications still needs to be determined, provided that the price of first-life batteries (FLBs) has been decreasing in recent years \cite{Vykhodtsev2022ASystems,Gruber2022ProfitabilityWinterization} and the refurbishment cost of SLBs is still high \cite{Madlener2017EconomicApplications,Wu2020DoesBatteries}. The economic viability of SLBs depends on various factors, including technical performance (i.e., battery degradation), which directly influences the service life of SLBs, refurbishment costs depending on the size of SLB packs, and potential revenues for all stakeholders (e.g., SLB sellers and buyers), which can vary between countries due to different policies. 
Finally, SLBs have lower power efficiency, heterogeneous electrical characteristics, and lower reliability compared to new batteries. Therefore, it is essential to design appropriate power electronics for SLB applications. Power electronics are the key energy conversion interfaces that can improve the performance of SLBs. However, designing power electronics for SLBs can be challenging, particularly for large SLB packs, due to significant cell-to-cell variation.

In a recent comprehensive review by Li et al. \cite{li_comprehensive_2022}, the SLB market was extensively evaluated from technical, economic, and policy perspectives. The authors discussed technical topics such as degradation models, the aging knee, and regrouping strategies, and summarized the economic performance of SLBs in practical applications. However, this review did not cover several key issues, including the specific application of hybrid physics-based and data-driven approaches for SLBs to characterize their degradation, the formulation of a grouping standard to balance refurbishment costs and SLB performance, an updated economic assessment of SLBs that highlights more recent results, and a comprehensive introduction to the advanced power electronics techniques that have been developed for SLBs. To this end, we aim to address all the above issues in this review. 

The remaining sections of this work are organized as follows. We first introduce various degradation mechanisms and models in Section \ref{sec:deg}. In Section \ref{sec:fastsc}, the process of dealing with SLBs, which mainly contains fast screening and regrouping, is presented, and the regrouping standards are first discussed. In Section \ref{sec:eco}, the economic benefits of SLBs in different applications are reviewed and classified from the aspects of understanding the potential benefits of SLBs, economic assessment strategies with critical assumptions, integration of battery degradation modeling, representative economic indicators, and the importance of market and government incentives.
In Section \ref{sec:power}, we discuss advanced power electronics technologies that enable the battery management system for SLBs. In Section \ref{sec:con}, the conclusions of this work are summarized. 

\section{Degradation Model}\label{sec:deg}

\begin{figure}[htbp] 
 \centering
   \includegraphics[width=\textwidth]{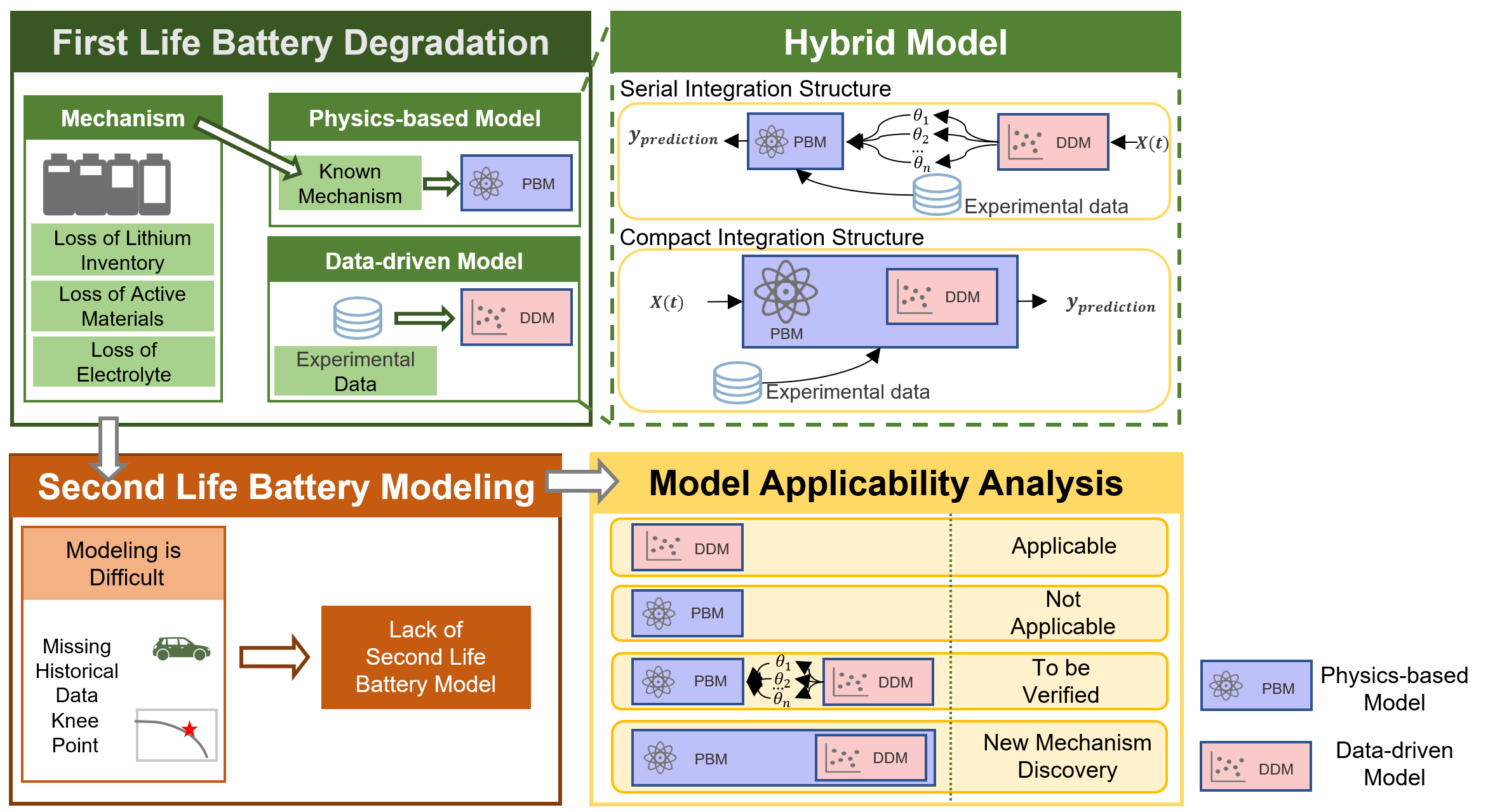}
       \caption{Schematic illustration of Section 2 on battery degradation models.}
  \label{fig:ch1}
\end{figure}

Main degradation mechanisms and models for FLBs have been extensively studied. As shown in Figure \ref{fig:ch1}, the degradation models primarily consist of three types: physics-based models (PBMs), data-driven models (DDMs), and hybrid models. Specifically, PBMs are explainable but may not be applicable when degradation pathways are poorly understood. In contrast, DDMs rely on high-quality data and bypass mechanisms but are not interpretable. 
The hybrid model leverages the advantages of both PBMs and DDMs and has the potential to establish an explainable model with physical laws that can be generalized.
The missing historical data over the first-life operation and the unknown knee point have posed noticeable challenges to accurately characterizing the degradation behaviors of SLBs. Currently, the degradation models particularly focusing on SLBs are still absent due to their unknown internal health status, which indicates the interplay of various aging mechanisms (e.g., SEI growth and lithium plating) and cannot be easily characterized using individual parameters (e.g., resistance or capacity). 
This section introduces the existing battery degradation models for FLBs and investigates their potential applicability in SLBs after appropriate modifications. With sufficient experimental data, hybrid models offer decent opportunities to reveal hidden mechanisms in SLBs while maintaining good interpretability.

\subsection{Introduction of degradation models}
\begin{figure}[htbp] 
 \centering
   \includegraphics[width=14 cm]{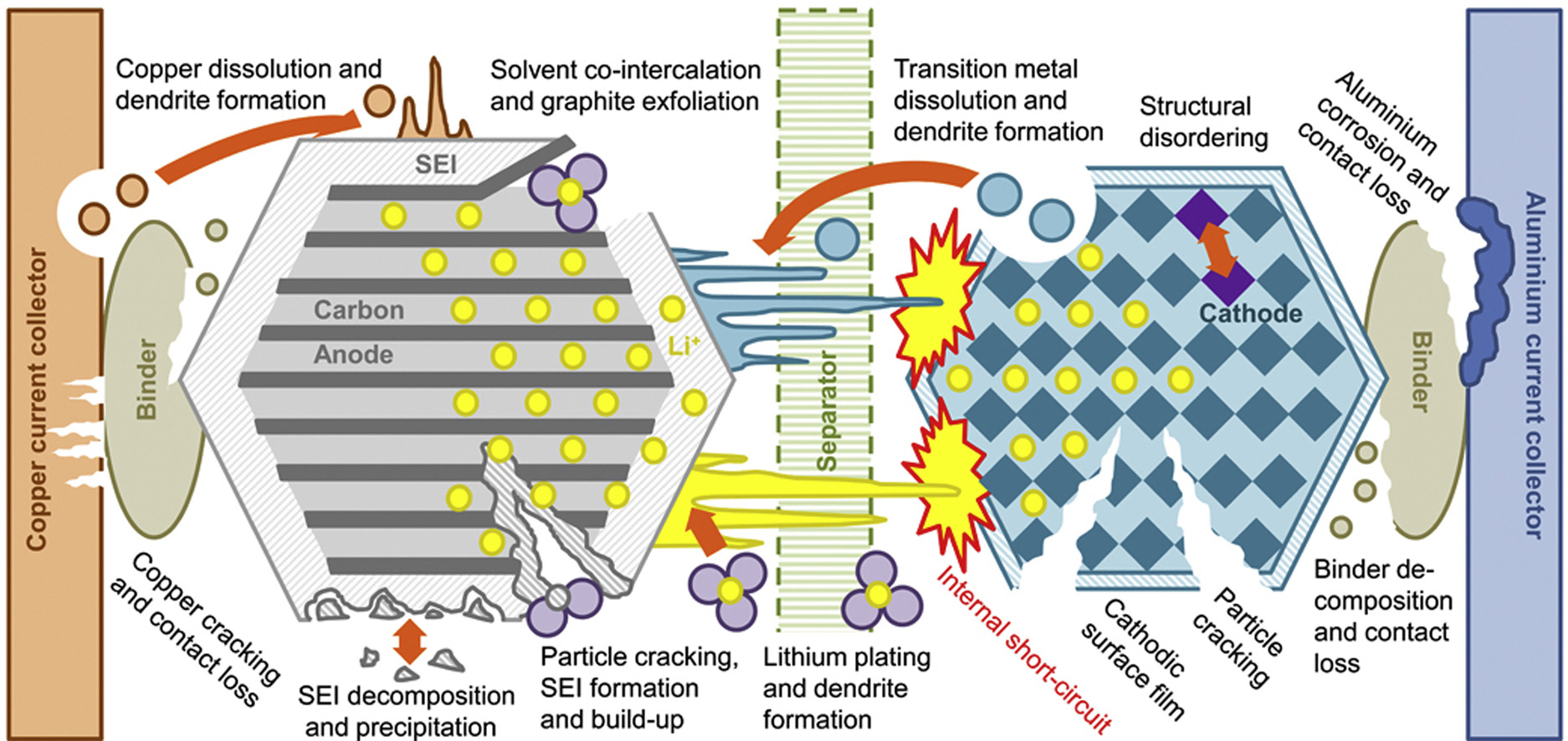}
  \caption{Various degradation mechanisms in lithium-ion batteries (Credit: \cite{Birkl2017DegradationCellsb}).}
  \label{fig:degraMech}
\end{figure}

The battery can be regarded as a dynamic electrochemical system including intricate internal reactions. The system will gradually deteriorate as the cycle number increases, caused by various mechanisms, such as physical stresses and chemical side reactions.
Figure \ref{fig:degraMech} provides an overview of the common battery degradation mechanisms, which can be classified into three main modes \cite{Birkl2017DegradationCellsb, Han2019ACycle}:
\begin{itemize}
    \item Loss of lithium inventory (LLI).
Lithium ions in the battery are consumed by surface film formation (SEI) growth, lithium plating, decomposition reactions, etc. These consumed lithium-ions can not transport between the positive electrode (PE) and the negative electrode (NE) anymore, leading to a reduction in capacity. Another manner in which lithium-ion loss occurs is through trapping in isolated active material particles.
    \item Loss of anode active material ($\rm {LAM_{NE}}$) and Loss of cathode active material ($\rm LAM_{PE}$). 
Particle cracking and loss of electric contact can lead to $\rm LAM_{NE}$, which increases the difficulty of lithium insertion into the active materials of NE. Particle cracking, structural disordering, and loss of electric contact can lead to $\rm LAM_{PE}$, making it difficult for lithium to enter the active materials of PE.
    \item Loss of electrolyte. 
Loss of electrolyte results from electrolyte consumption due to side reactions like SEI formation and thickening, as well as electrolyte decomposition caused by high potential or high temperature. The loss of electrolyte will lead to an increase in resistance and affects the capacity and power capability of the battery. Significant electrolyte loss may even result in a considerate capacity drop  \cite{Han2019ACycle}.   
\end{itemize}

As shown in Figure \ref{fig:degraMech}, the intricate internal changes inside the battery, involving various electrochemical side reactions in the anode, electrolyte, cathode, and current collectors, make the battery degradation modeling a complex problem. In addition, the operation condition also strongly affects the degradation rate of the battery (e.g., temperatures and current amplitudes) and may induce different aging mechanisms. These factors make it challenging to quantify battery degradation under different working conditions. Consequently, degradation modeling has always been an active research area, with significant studies available in the literature \cite{Hu2020BatteryPrognostics}. Various kinds of degradation models can be summarized as follows:     
\begin{itemize}
\item PBMs.
Battery degradation is a complex problem caused by multiple electrochemical side reactions in the anode, electrolyte, cathode, and current collectors. To trace the degradation pathway based on mathematical models, PBMs explore the fundamental processes in the battery. 
The SEI formation is one of the significant factors for battery degradation, which highly accounts for the initial and linear capacity decrease \cite{Han2019ACycle, Stamps2005AnalysisBattery}. Since lithium-ion is a constitution of the SEI layer, its formation will lead to the LLI. Besides, lithium-ion moving through the SEI layer will experience higher ionic resistance. Efforts can be seen in the literature to describe the formation of the SEI layer from macroscopic levels \cite{Roder2019DirectSystems} and molecular levels, such as kinetic Monte Carlo simulation \cite{Methekar2011KineticFormation}, molecular dynamics \cite{Leung2010AbAnodes}, and density functional theory \cite{Leung2013ElectronicBatteries}.
Lithium plating is another important mechanism contributing to degradation by consuming lithium-ions and increasing impedance. Considering both lithium plating and SEI growth, Yang et al. \cite{Yang2017ModelingAging} built an electrochemical-thermal model and explained that SEI growth should be responsible for the linear degradation and lithium plating causes the nonlinear aging trends in the second stage. It was found that SEI growth can lead to the anode porosity drop and thus lower lithium deposition potential in the charging process. Lithium plating further lowers anode porosity and significantly accelerates the battery degradation process caused by a positive feedback loop, thereby finally leading to the end-of-life of batteries. 
Particle fracture is also a typical physical process inside batteries, which exposes an extra surface area for further SEI growth and causes LLI. O'Kane et al. \cite{OKane2022lithium-ionIt} established a PBM that incorporates these mechanisms and showed that LLI significantly increases when particle cracking is considered.
    
PBMs offer several natural advantages. 
Firstly, degradation models provide valuable insights into battery aging owing to the well-understood internal processes. For instance, Yang et al. \cite{Yang2017ModelingAging} demonstrated that an internal mechanism such as SEI growth or lithium plating could dominate each degradation phase. 
Secondly, the knowledge gained from PBMs can be leveraged for future engineering design. For example, Han et al. \cite{Han2019ACycle} demonstrated that the internal side reactions were directly influenced by the battery design. By integrating physical models into the design optimization, it is possible to mitigate internal side reactions and enhance the battery lifespan.
Thirdly, this kind of method can well extrapolate to other cases if the dominant aging mechanism remains the same. However, degradation characterization for SLBs can still be challenging as the electrochemical states/parameters are difficult to measure and estimate \cite{UlrikeKrewer2018Review-DynamicPerspective}, especially when the historical data is missing, leaving the degradation pathway over the first-life application largely unknown. 

\item DDMs.
DDMs take advantage of historical data to train a model to predict degradation trends. Machine learning frameworks involving models like Naive Bayes \cite{Ng2014ABattery}, Support Vector Regression \cite{Wang2014PrognosticsRegression}, Gaussian Process Regression \cite{Richardson2017GaussianHealth, Li2016RemainingMixture}, and Neural Networks \cite{Zhang2018LongBatteries, Liu2015AnEstimation} were well practiced. 
Severson et al. \cite{Severson2019} collected a comprehensive dataset of 124 cells and built a feature-based linear model, which can predict the remaining life using the first 100 cycles with an error of 9.1\%. Using the same data, Hsu et al. \cite{Hsu2022DeepOnly} developed a deep neural network that can predict the remaining life by the first cycle in a lower test error of 6.46\%, indicating the promising performance of the extracted features.

DDMs have shown great potential to predict battery degradation based on available high-quality data. 
Such datasets, however, are challenging to acquire in the academic community because of limited resources. 
Opportunities exist in leveraging enormous data from the on-road EVs, which can be used to learn the mission profiles of batteries in their first life. The operational data in EVs on the road can be leveraged as prior information about how the onboard batteries age in their first life. 
The main challenge of utilizing field data is the rigorous propagation of uncertainty caused by uncontrolled operating conditions and low-quality data. Bayesian methods for parameter estimation and prediction are a promising approach since they provide simultaneous prediction and observation uncertainty with realistic confidence bounds \cite{sulzer_challenge_2021}. 
Besides, DDMs require parameter calibration or training for new batteries, and the extracted features can be challenging to interpret, even with high prediction accuracy.

In addition, the empirical model is also established based on data along with engineering experience. This kind of model produces a regression model that represents the degradation trend considering several important factors, such as time, temperature, C-rate  \cite{Wright2002Calendar-Batteries, Bloom2001AnCells}, internal resistance \cite{Tseng2015RegressionBatteries}, battery discharge curve \cite{Lu2017ModelingLife}, etc. Empirical models are relatively simple and are often investigated in literature to predict battery degradation and evaluate its remaining value \cite{Seger2022AAnalysis}, as will be discussed in section \ref{sec:eco}. 

\item Hybrid models.
Hybrid approaches aim to combine the strengths of both PBMs and DDMs to better predict battery lifetime. The reviews in \cite{Aykol2021PerspectiveCombiningLifetime, guo_review_2022} discussed various integration structures for predicting battery lifetime, which can be broadly divided into serial integration structures and compact integration structures, as shown in Figure \ref{fig:hybrid}.
One integration approach is residual learning, where DDM can learn the residuals between the PBM outputs and the experimental outputs, as shown in structure A1 of Figure \ref{fig:hybrid}. Park et al. \cite{saehong_park_hybrid_2017} integrated a recurrent neural network with the single particle model to recover model-measurement mismatch and improve the prediction accuracy of the output voltage.
Another approach is transfer learning, where the DDM learns from the data generated by a PBM and a few experimental data, as shown in structure A2 of Figure \ref{fig:hybrid}. In this structure, the PBM performs as the data generator to reduce the dependency of a DDM on experimental data. Li et al. \cite{li_physics-informed_2021} used an experimentally verified pseudo 2D (P2D) model to generate much data to train a deep neural network for estimating battery states. 
Parameter identification is also a typical hybrid structure, where the DDM estimates the best parameters for a PBM to explain the experimental data, as shown in structure A3. Reports of such an approach can be widely seen in the literature. A recent work by Li et al. \cite{li_data-driven_2022} applied the cuckoo search algorithm to identify 26 parameters for the P2D model through a two-step identification process and a multi-objective fitness function. 

\begin{figure}[htbp] 
 \centering
   \includegraphics[width=16 cm]{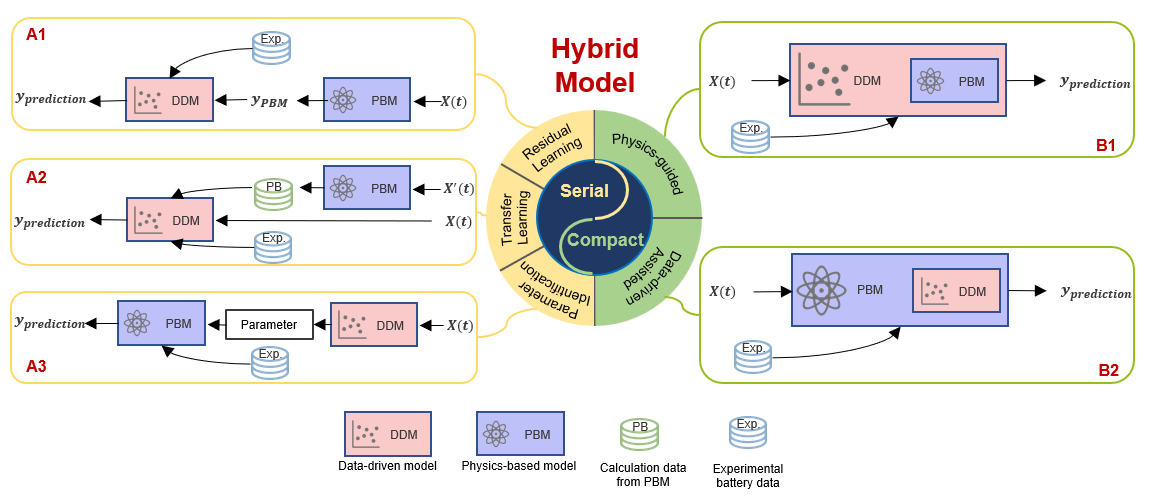}
  \caption{Battery hybrid degradation model. Serial structures: A1 - residual learning, A2 - transfer learning, A3 - parameter identification; compact structures: B1 - physics-guided model, B2 - data-driven assisted model (recreated from \cite{aykol_perspectivecombining_2021}).
  }
  \label{fig:hybrid}
\end{figure}

Besides the three serial integration structures, PBMs and DDMs can be integrated into compact structures, as shown in structures B1 and B2. Specifically, structure B1 is typically accomplished either by a physics-guided loss function and/or model architecture on the basis of the DDM, while in structure B2 part of the physical model (usually costly or inaccurate) is replaced with a DDM. The PINN is an example of structure B2, the essence of which is to solve partial differential equations (PDEs) using deep learning trained by automatic differentiation \cite{karniadakis_physics-informed_2021}. 

By incorporating information on physical laws, boundary conditions, and initial conditions, PINN can effectively solve PDEs, and it offers several advantages over traditional methods, such as being meshless and able to solve PDEs for all possible parameters, fast inference, and the ability to solve inverse problems. 
Zubov et al. \cite{zubov_neuralpde_nodate} conducted an initial investigation of the use of PINN to solve the single particle model and the reduced-order P2D model. They obtained satisfactory results for most variables but experienced large errors for certain state variables, such as electrolyte potential. However, in light of the rapid advancements in this field and the emergence of more effective training methods, it is promising to see the PINN solution for a full P2D model across a broad range of parameters. As the battery degrades, the parameters of the governing equations will change as well, which can make the degradation model more complex. Since no related research has been found yet to exploit the adaptive PINN as degradation models, this can be a future research direction in terms of organically leveraging both explainable degradation mechanisms and learning capabilities for high-dimensional correlations with limited degradation data.
\end{itemize}

\subsection{Degradation models for SLBs}
While substantial degradation models have been provided in the literature, there is still a dearth of studies focused on degradation models specifically designed for SLBs. This is a more challenging task for two reasons: firstly, the lack of historical data for first-life applications, and secondly, the uncertainty surrounding the ``knee point''. This section provides a comprehensive overview of these issues.

\subsubsection{First-life operation data}
The historical operation data of SLBs over their first-life applications significantly affect how the battery ages in second-life applications. For example, the fast charging in a cold environment will cause significant lithium plating, which leads to capacity loss and even takes the risk of thermal runaway caused by lithium dendrites penetrating the separator \cite{Hu2022AApplications}. 
The research in \cite{Waldmann2017ElectrochemicalApplications} explored the effects of working conditions on battery degradation throughout its entire lifetime. Through experimentation on multiple 18650-type cells at two different temperatures (0$^{\circ}$C and 45$^{\circ}$C), it was observed that lithium plating occurred at 0$^{\circ}$C/0.5 C-rate, while SEI growth and $\rm LAM_{NE}$ occurred at 45$^{\circ}$C/0.5 C-rate. These findings suggest that the internal degradation mechanism of the battery was significantly impacted by the operating condition. 

Although historical data is critical for understanding the degradation pathway of SLBs, it is often unavailable after they are disassembled from EV battery packs. Note that the simple parameters (e.g., remaining capacity and internal resistance) are not enough to well describe how the battery has aged. For example, two battery cells, both with 80\% remaining capacities, may have totally different remaining lifetimes in their second-life applications due to different internal statuses after the first life. A potential research direction to address this issue is to develop simple testing protocols, which can be used to provide informative aging information on SLBs by estimating and assessing the key parameters related to various aging mechanisms.

\subsubsection{Knee point}
\begin{figure}[htbp] 
 \centering
   \includegraphics[width=10 cm]{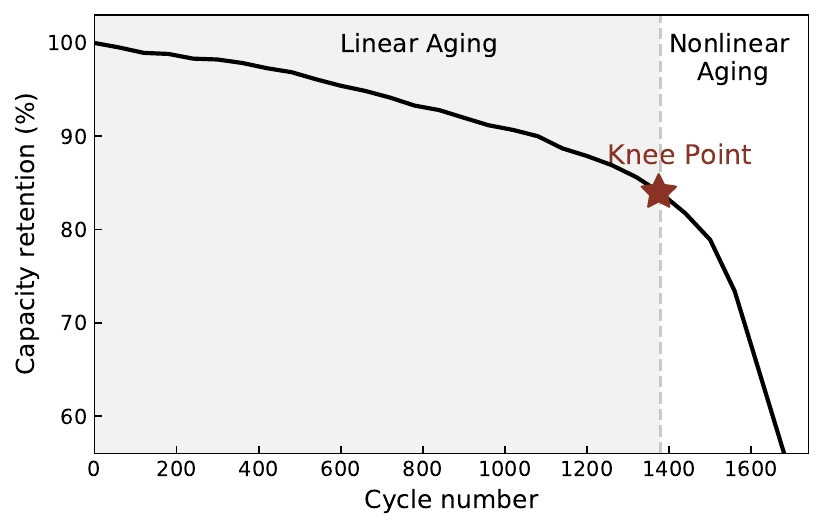}
  \caption{The knee point during battery degradation}
  \label{fig:knee}
\end{figure}

The degradation curve of a cell can be drawn as the change in capacity over cycle number, as depicted in Figure \ref{fig:knee}. During the early stages of operation, the curve will exhibit a linear trend. However, beyond a certain point, the performance of the cell will experience a sharp decline, and the transition point from a mostly linear region to a highly non-linear region is commonly referred to as the knee point.
After the knee point, the battery capacity will rapidly decrease, and severe safety issues may arise. The internal mechanisms responsible for the knee are convoluted and not yet fully understood.

The review by Attia et al. \cite{Attia2022ReviewKneesTrajectories} covered 303 cells from 17 datasets and calculated the relationship between the knee point and the EOL point (80\% of nominal capacity), which can be represented by Eq. \eqref{eq:eolknee}.

\begin{equation}
    \label{eq:eolknee}
    y = 0.984x + 93.918 \ (R^2 = 0.874)
\end{equation}
where $x$ represents the cycle number to the knee point and $y$ is the cycle number to EOL. The equation has shown a clear linear relationship and demonstrated that the occurrence of the knee is always accompanied by the EOL point, i.e., the retired batteries take higher risks of encountering the knee point. For safe SLB usage, more light should be shed on accurate knee point prediction. However, a comprehensive framework with accurate models for predicting knees of battery cells with various chemistries is still missing. In \cite{Attia2022ReviewKneesTrajectories}, the failure modes leading to knees were categorized into six pathways: lithium plating, electrode saturation, percolation-limited connectivity, resistance growth, additive depletion, and mechanical deformation.  Meanwhile, the knees can be classified into one of three “internal state trajectories”: snowball, hidden, and threshold. Because each mechanism is represented by a different pair of pathway-trajectory, the problem of modeling and predicting knees is extremely challenging.


\subsubsection{Applicability of the existing degradation models for SLBs}

As mentioned above, it is still unclear whether the degradation models mentioned above for FLBs are still applicable to SLBs. 
The DDM, due to its nature of fitting an approximate function and bypassing mechanisms, could still perform well if specific lab test datasets are established for SLBs. Some research like \cite{Severson2019Data-drivenDegradation} can be conducted for SLBs to build degradation models by data-driven approaches. Besides, such methods will greatly benefit from open-source datasets, hence it is recommended to make valuable datasets open-source to provide abundant data for degradation modeling in SLBs (similar to current famous FLB datasets, including but not limited to NASA datasets \cite{B.Saha2007BatterySet,Bole2014AdaptationUse}, CALCE datasets \cite{UniversityofMaryland2014CALCEDatasets}, Stanford/Toyota datasets \cite{Severson2019Data-drivenDegradation, Attia2020Closed-loopLearning}, Oxford battery degradation datasets \cite{Birkl2017Oxford1}, Panasonic 18650PF Li-ion battery datasets \cite{Kollmeyer2018PanasonicData}, and NCM/NCA datasets \cite{zhu_data-driven_2022}).

For PBMs, the prerequisite is understanding the dominant mechanism responsible for the degradation pathway over the first life.
Therefore, a simple testing protocol, which can effectively distinguish the degradation mechanisms for SLBs, would be useful. Some modifications are required to enable the implementation of PBMs in SLBs.
As for hybrid models, it remains unclear whether structures A1-A3 can accurately capture the SLB degradation trend, as there is a lack of research in this area, while the use of new SLB experimental data could potentially mitigate prediction errors caused by unknown internal mechanisms.
To reveal unknown mechanisms in SLBs, a hybrid model with the B2 structure (e.g., PINN) can be applied to solve the inverse problem by leveraging observation data.
Finegan et al. recently outlined PINN approaches to extracting physical insights from cell failure events, which can improve the safety of batteries \cite{Finegan2021PerspectiveSafety}. Thus, hybrid models show promise as a solution for better understanding the internal mechanisms of SLB degradation.

\section{Fast Screening and Regrouping}\label{sec:fastsc}

\begin{figure}[htbp] 
 \centering
   \includegraphics[width=15 cm]{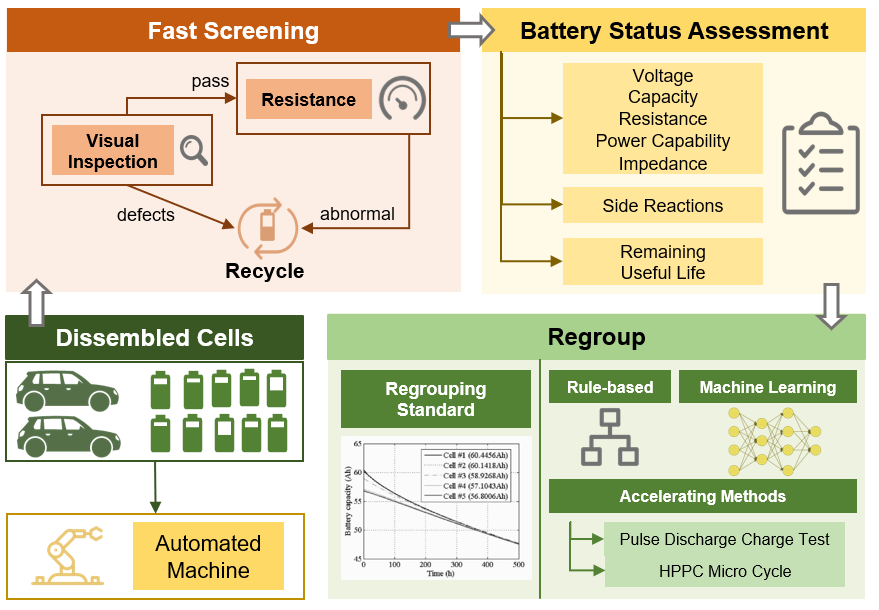}
  \caption{Schematic illustration of Section 3 on battery fast screening and  regrouping.}
  \label{fig:ch2}
\end{figure}

Generally, the retired batteries disassembled from the battery pack will experience fast screening and regrouping before they can be reused in second-life applications. As shown in Figure \ref{fig:ch2}, we first introduce the fast screening, which includes visual inspection and fast parameter check. Those batteries that pass the fast screening are then subjected to detailed measurements, through which informative features that indicate the performance of SLBs are extracted. The batteries are finally classified into groups based on the extracted indicators by rule-based or machine learning methods. Several potential methods to accelerate the entire process are discussed, and the regrouping standard for SLB cells is investigated.

\subsection{Fast screening}
To efficiently refurbish a large number of heterogeneous battery cells/modules/packs, a fast screening procedure is urgently required for SLBs. This procedure consists of visual inspection and parameter assessment. The shape, size, and deformation of battery cells will be initially evaluated in the visual inspection, and cells with qualified appearances undergo testing to obtain critical health-oriented parameters or features. 
For example, internal resistance can be fast measured, and those batteries with normal resistance can go to the next procedure, while the abnormal batteries will be directly recycled \cite{Garg2020}. 
Note that individual parameters such as resistance cannot indicate the overall performance of SLBs. Other battery parameters, including capacity, voltage, power capability, impedance, and temperature, can be synergized to assess battery conditions. Table \ref{tab:para} lists some key parameters and the corresponding measurement methods. In order to reflect the internal status of SLBs and the remaining life expectancy, Lai et al. \cite{Lai2021} proposed a way to evaluate SLBs more comprehensively, and a heating impedance method is applied to obtain direct current (DC) resistance which can indicate the lithium plating severity.

Due to the high volume of EVs currently in service and the limited lifetime of onboard LIBs, there will be a growing demand for the disassembly and screening of SLBs. This presents an opportunity to develop a scalable automated machine that incorporates robotic disassembly and automatic screening.  
Although some literature exists on the subject, such as the review by Meng et al. \cite{Meng2022IntelligentOverview}, which highlights the role of AI techniques in improving intelligent disassembly for LIBs, the cases discussed mostly come from academia and lack real-world verification. Therefore, there is a need to focus more on industrial practices, which are more closely relevant to practical applications.

\begin{table}[htbp]
  \centering
  \caption{Some key parameters and test methods.}
  \begin{threeparttable}
    \begin{tabular}{p{7.39em}ll}
    \toprule
    Parameter & Test Method &         \\
    \midrule
    Capacity & CC-CV test \cite{Martinez-Laserna2018TechnicalPerspective} &         \\
    Voltage  & quasi-OCV vs SOC test \cite{Martinez-Laserna2018TechnicalPerspective} &         \\
    DC resistance & Interal resistance tester \cite{Garg2020},  Hybird pulse power characterisation \cite{Martinez-Laserna2018TechnicalPerspective}  &  \\
    Power capability & Hybird pulse power characterisation \cite{Martinez-Laserna2018TechnicalPerspective} &   \\
    Impedance & Electrochemical impedance spectroscopy \cite{Li2020} &   \\
    Temperature & Surface sensors \cite{Garg2020} & \\
    \bottomrule
    \end{tabular}%
  \end{threeparttable}
  \label{tab:para}%
\end{table}%

\subsection{Regrouping}
\subsubsection{Regrouping methods}
Batteries can be potentially classified by their parameters and categorized into different groups by human-made rules. For example, according to the remaining voltage values and fast/slow charge/discharge performance, the study in \cite{Schneider2014ClassificationPrototypes} classified NiMH and Li-ion batteries based on the remaining voltage, and retired batteries were then repurposed for use in a portable power supply and illuminator. Given that the performance of retired batteries is highly related to the working conditions in second-life applications, the effective capacity, i.e., the capacity of batteries at a specific discharge rate and temperature, was proposed in \cite{Li2017} to tackle this issue. This allows retired batteries to be effectively regrouped to meet application requirements. 
These works achieve classification by following specific artificial rules requiring expert experience and manual classification. To automatically regroup heterogeneous batteries, some authors proposed using machine learning approaches, including but not limited to the K-means algorithm \cite{Jiang2014ResearchModel,Celebi2013AAlgorithm}, self-organized network \cite{Garg2020}, and support vector machine regression analysis \cite{Choubin2019AnMachines}. Machine learning approaches can synergize the measured parameters comprehensively and achieve an automatic classification. However, the interpretability of these methods needs to be improved to account for the intertwined degradation mechanisms in SLBs. 
The work presented in \cite{Lai2021} developed a more comprehensive regrouping method for SLBs. This method utilized a one-dimensional classification algorithm to evaluate side reactions (e.g., lithium plating) of batteries. The batteries with the same side reaction characteristics are further classified according to different applications for SLBs, including energy scenarios, power scenarios, and energy-power scenarios. In an energy scenario, priority was given to capacity and remaining life, whereas in a power scenario, the main concerns were internal resistance and remaining life. For a power-energy scenario, capacity, internal resistance, and remaining life were all considered.

It is essential to accelerate the classification efficiency for reduced refurbishment cost, which directly determines the economic performance of SLBs \cite{song_economy_2019}. 
One class of acceleration approach is to apply simple test protocols and avoid the cycle test for SLBs. For instance, the research in \cite{Zhou2020} classified large amounts of retired batteries effectively by combining a fast pulse test with an improved bisecting K-means algorithm. The classification efficiency was improved without impairing performance by using pulse test curves as features in addition to typical voltage, capacity, and resistance (U/Q/R) features. This method reduced the time cost from 5 hours to 2 minutes. 
The second class of methods searches for improvement in parameter measurements. For example, the work by \cite{Muhammad2019} proposed a method that uses offline hybrid pulse power characterization micro-test data to benchmark cells' response over an SoC range. The recovering time after the pulse tests reflects the strength of the battery, with stronger batteries recovering more quickly than weaker ones, indicating higher power capability and capacity. The proposed technique efficiently distinguish different batteries in 80 seconds. 


\subsubsection{Regrouping standards}
Although repurposing batteries has been investigated in the existing work, the regrouping standards, indicating the tolerance of cell-to-cell variation within a battery pack, remain unclear. It is essential to establish a clear and reasonable standard that can serve as a guideline for repurposing batteries to enhance the performance of SLBs, as it is still unclear how cell-to-cell variation impacts pack-level performance (e.g., efficiency and safety). 
An accurate quantitative standard of regrouping batteries should be established by incorporating the self-balancing mechanism among parallel-connected battery cells, meaning that the progression of cell-to-cell variation regarding capacities will stabilize under the convex or linear degradation curves \cite{Song2021}. Therefore, the cell-to-cell variation is tolerable to some extent when regrouping batteries. This will benefit battery classification because cells with mild variation can be reasonably clustered into the same group. However, the allowable variation should be further quantified for different battery designs and operation scenarios. 
In \cite{Song2022}, cell-to-cell variation within parallel-connected cells under different cooling structures was analyzed, and it found that the round cooling structure with relatively uniform cooling conditions can significantly reduce the cell-to-cell variation compared with the sequential structure with a non-uniform cooling structure. However, there is a trade-off between system cost and suppressing cell-to-cell variation, as uniform cooling structures are generally complex and costly. 
Given the above analysis, the stabilized cell-to-cell variation can be adopted as the upper limit of parameter variation for SLBs.

\section{Economic Analysis}\label{sec:eco}

\begin{figure}[htbp] 
 \centering
   \includegraphics[width=\textwidth]{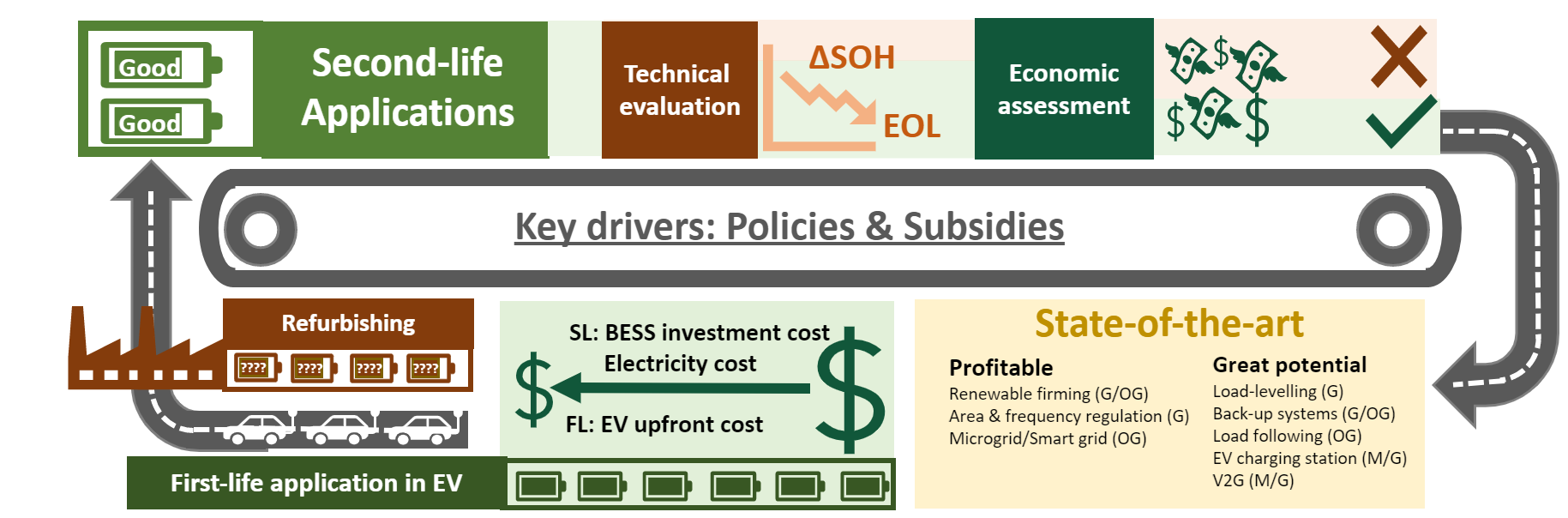}
  \caption{SLBs benefit assessment strategies and the current state-of-the-art.}
  \label{fig:economic structure}
\end{figure}

This section reviews and analyzes the economic benefits of reusing batteries retired from EVs in different applications. Despite numerous studies on the subject, the profitability of SLB applications remains unclear due to the uncertainties and assumptions underlying the point of interest \cite{Zhu2021End-of-lifeBatteries}. Thus, it is crucial to identify the deterministic factors involved in the assessment and understand the advantages and disadvantages of different assessment schemes.
Furthermore, the profitability of SLBs is heavily influenced by government incentives and policies as well as the service life of SLBs in their second-life applications \cite{Zhu2021End-of-lifeBatteries}, as illustrated in Figure \ref{fig:economic structure}. Before using SLBs in second-life applications, the economy of SLBs needs to be studied, and the profitable applications should be further promoted. By reusing batteries, we aim to reduce the costs for both first-life and second-life applications, thereby promoting the circular economy of LIBs.

\subsection{State-of-the-art}
The first techno-economic analysis for SLBs was presented by the U.S. Advanced Battery Consortium \cite{NaumPinskyUSABCProgramManager1998ElectricStudy}. Later, Cready et al. \cite{osti_809607} estimated the investment costs of various SLB applications, considering the costs associated with acquisition, transportation, battery testing, and refurbishment. Transmission support, light commercial load following, residential load following, as well as distributed node telecommunications backup power were considered note-worthy and might be profitable for reusing batteries at that time.
Overall, SLB applications can be categorized into three types: on-grid, off-grid, and mobile applications. Presently, on-grid applications are more profitable than off-grid ones \cite{Sun2020EconomicChina}. Specifically, for on-grid applications, SLBs are beneficial when applied to renewable firming, area and frequency regulation, and peak shaving, as summarized in Table 
Furthermore, using SLBs in micro-grid and smart-grid applications has been proven to be feasible in work \cite{Debnath2016GridableGrid}. Although few studies have explored SLBs for mobile applications, fast charging-related applications have shown great potential \cite{Hossain2019,Martinez-Laserna2018}.
Most existing studies compare the economic performance of energy storage system (ESS) designs with or without SLBs in a single application \cite{Kamath2020EconomicStorage,Gur2018TheAnalysis,Assuncao2016TechnicalEnergy}. Due to inconsistencies in the methodology used for system modeling and economic assessment across various studies, a comparison of the economic performances of SLBs in different applications is still missing.

\begin{table}[htbp]\scriptsize  
  \centering
  \caption{A summary of the economic performances and assessment methodologies of SLBs in various applications.\\  Green: Profitable, white: might be profitable, red: not profitable.}
  \begin{threeparttable}
    \begin{tabular}{p{1.5em}p{8em}p{7em}p{12em}p{8em}p{3em}}
    \toprule
    &Application&Economic model&Degradation model&SLBs cost consideration&Discount rate \\
    \hline
    \multicolumn{6}{p{12em}}{\textbf{On-grid applications}} \\
    \midrule
    \rowcolor[HTML]{CBE684}\cite{Sun2020EconomicChina}&Peak-shaving&NPV, cost-benefit model&Constant decline rate& 15\%-40\% of the FLB cost&8\%\\
    \rowcolor[HTML]{CBE684}\cite{Assuncao2016TechnicalEnergy}&PV-BESS&NPV, payback period &Semi-empirical model \cite{Xu2018ModelingAssessment}&Nissan Leaf: 1496 \$/cell, Citroen Co: 2175 \$/cell&3\%, 5\%,7\%\\
    \rowcolor[HTML]{CBE684}\cite{Neubauer2012AValue} &Uninterrupted power supply& Cost-benefit model&Constant throughput degradation model&38-132 \$/kWh&10\% \\
    \rowcolor[HTML]{CBE684}\cite{Saez-De-Ibarra2015SecondService} & Demand responds & COE& DOD-based aging model, rain-flow cycle counting algorithm&-&-  \\
    \rowcolor[HTML]{CBE684}\cite{Mathews2020TechnoeconomicAging}&PV-BESS&Cost-benefit model&Semi-empirical data-based model \cite{Xu2018ModelingAssessment}&104.5 \$/kWh&7\%\\
     \rowcolor[HTML]{CBE684}\cite{Kamath2020EvaluatingApplications}& Peak shaving &NPC,LCOE&Lifetime throughput model\cite{Debnath2016GridableGrid}&65 \$/kWh\cite{osti_809607}&6.9\%\\
    \rowcolor[HTML]{CBE684}\cite{Wolfs2011AnSupport}&Grid-support&NPV&Weighted average Amp-throughput model&74 AUD/kWh&4.5\%\\
    \rowcolor[HTML]{CBE684}\cite{Kamath2020EconomicStorage}&Fast charging energy storage&LCOE&-&95 \$/kWh& 6.7\%\\

    \cite{Gur2018TheAnalysis}&PV-BESS&NPV&-&120-450 \texteuro/kWh&1\%-10\%\\
    \cite{Neubauer2011TheApplications}&Utility-scale applications&NPV&Constant throughput degradation model&30\% of the FLB cost&10\%\\
    \cite{Heymans2014EconomicLoad-levelling}&  Load-leveling &NPV&- &55-132 \$/kWh&1\%\\
    \cite{Wu2020DoesBatteries}&Energy arbitrage&Cost-benefit model&Gaussian process regression model\cite{Yang2018ACurve}&112-128 \$/kWh&2\%-8\%\\
    \cite{Xu2021StudyMarkets}&Regulation market&NPV&DOD-based model, rain-flow cycle counting algorithm&126 \$/kWh&5\%\\
    \cite{Fallah2022HowScenarios}&Energy, non-energy service&Cost-benefit model&DOD-based model&350 \texteuro/kWh&7\%\\
    \cite{Rallo2020lithium-ionCases}&Energy arbitrage, peak shaving&Annual saving, return-on-investment& Equivalent circuit model &50 \texteuro/kWh&-\\
    \rowcolor[HTML]{FFB995}\cite{Steckel2021ApplyingSystems}&Utility-scale application&LCOE, LCOS&Degradation  rate at 1\%-3\% per year&64.3\%-78.9\% of the FLB cost&-\\  
    \rowcolor[HTML]{FFB995}\cite{Song2019EconomyScenarios}&Wind farm&Equivalent price model&Semi-empirical model \cite{Song2014Multi-objectiveVehicles}&100 \$/kWh&-\\  
    \hline
    \multicolumn{6}{p{37em}}{\textbf{Off-grid applications}} \\
    \hline
    \rowcolor[HTML]{CBE684}\cite{Debnath2016GridableGrid}&Smart grid generation side asset management&Capital investment, COE&-&100 \$/kWh&5-10\%\\
    \rowcolor[HTML]{CBE684}\cite{Alimisis2013EvaluationPower}&Hybrid wind farm&Revenue, annualized cost&Dynamic degradation model&86 \texteuro/kWh&8.42\%\\
    \rowcolor[HTML]{CBE684}\cite{Bai2019EconomicChina}&PV-BESS&NPC,LCOE&-&50 CNY/kWh&-\\
    \cite{CanalsCasals2019ReusedServices}&Smart-grid&Cost, revenue&Equivalent circuit model &38-140 \texteuro/kWh&-\\
    \rowcolor[HTML]{FFB995}\cite{Bhatt2022OptimalSystem}&Microgrid&NPV,COE&Modified kinetic battery model&-&-\\
    \hline
    \multicolumn{6}{p{37em}}{\textbf{Mobile applications}} \\
    \hline
    
    \rowcolor[HTML]{CBE684}\cite{Funke2020FastCities}&Fast charging station& Total cost of ownership, NPV &-&100 \$/kWh&5\%\\  
    \cite{Graber2020BatteryStations}&EV charging station& COE &DOD-based linear model &-&4\%\\  
    \bottomrule
    \end{tabular}%
  \end{threeparttable}
  \label{tab:eco}%
\end{table}%

\subsection{Economic benefits of SLBs}

\subsubsection{First-life benefits}
Repurposing EV batteries in stationary applications can lower the upfront cost of EVs and the cost of FLBs \cite{Neubauer2011TheApplications}. The work by Debnath et al. \cite{Debnath2014QuantifyingGrid} evaluated the effect of reusing batteries in the smart grid on the price of FLBs and the benefits for EV owners. The results indicated that the revenue generated by SLBs compensates for 19.56\% of the FLB's purchase cost. A similar conclusion was obtained in \cite{Wu2020DoesBatteries} that the profit of repurposing SLBs for stationary energy storage applications would reduce the EV's upfront cost by 14.3\% to 36.3\%. 
 
\subsubsection{Second-life benefits}
Compared with a fresh battery energy storage system (BESS), the investment cost can be reduced by up to 60\% when SLBs are integrated with photovoltaic (PV) panels \cite{Mathews2020TechnoeconomicAging}, 70\% when SLBs are used as energy backup sources \cite{Debnath2016GridableGrid}, and 73.62\% when SLBs are in a grid-connected renewable energy system \cite{Bhatt2022OptimalSystem}.

Using SLBs in a BESS can also reduce the cost of energy (COE), as proved in the studies considering local tariffs in different countries, such as Portugal \cite{Assuncao2016TechnicalEnergy}, Germany \cite{Madlener2017EconomicApplications}, and China \cite{Bai2019EconomicChina}. 
The study by \cite{MirzaeiOmrani2019EconomicSectors} has demonstrated that using SLBs for load leveling reduces the electricity bill by up to 39.7\%  because integrating BESS with renewable generation facilities results in a higher self-consumption rate \cite{Kamath2020EconomicStorage}. However, the efficacy of COE reduction from SLB integration depends on the scale of the renewable power plant. SLBs are more suited for large-to-medium-scale applications in China in terms of promoting a self-consumption rate compared to residential applications \cite{Bai2019EconomicChina}. Similar conclusions are found in \cite{Gur2018TheAnalysis}, which suggested that SLBs would be crucial for utility-scale BESS, also known as front-of-the-meter battery storage \cite{InternationalRenewableEnergyAgency2019UTILITY-SCALEBRIEF}, due to more intense energy demand. Overall, these studies highlight that SLBs would be an alternative to FLBs in various applications, especially supply-side management, as they help lower both the capital investment of ESS and COE.
 
\subsection{System modelling}
 
\subsubsection{Economic models}
Various economic models can be found in existing studies to assess the economic benefits of SLBs: such as the net present value (NPV) model, which estimates the net present worth of the SLB project, including time-varying costs and revenue; the levelized cost of electricity (LCOE) model, which stands for the average cost of electricity considering time-varying investment costs and energy production; the levelized cost of storage (LCOS) model, which is similar to the LCOE model but only includes energy storage-related costs in calculations; and the cost-benefit model, which represents the ratio between the investment cost and the economic benefit obtained from SLBs. In the aforementioned economic models, the capital recovery factor, calculated by the discount rate and the estimated lifetime of the project, is used to convert future cash flows into present values.

The NPV model considers the present value of cash inflows and outflows throughout the project and is widely used in capital budgeting and investment planning \cite{2022NetIt}. In the case of SLB applications, a positive NPV indicates profitability. For example, Bai et al. \cite{Bai2019EconomicChina} evaluated the NPV of distributed solar photovoltaic systems in China and found that reusing batteries is profitable nationally for both the industrial and commercial sectors. However, Gur et al. \cite{Gur2018TheAnalysis} reached the opposite conclusion when analyzing the European electricity market.

The LCOE model can provide an explicit comparison between the economic performance of SLBs and FLBs, with the flexibility to be reformulated to incorporate other forms of expenditure. For example, in fast charging applications, the LCOE could be reduced by 12\%-14\% using SLBs, although this result is dependent on several factors, including the cost and lifetime of the SLBs, their efficiency, and the discount rate \cite{Kamath2020EconomicStorage}. Steckel et al. \cite{Steckel2021} proposed LCOS and found that the LCOS of SLB-ESS (i.e., 234-278 \$/MWh) is higher than that of FLB-ESS (i.e., 211 \$/MWh), but the upfront costs for SLB-ESS are only 64.3\%-78.9\% of those of FLB-ESS.

The cost-benefit model is useful for analyzing the expenditures and gains of using SLBs in ESS. Studies by Fallah et al. \cite{Fallah2021End-of-LifeModelling} and Rallo et al. \cite{Rallo2020lithium-ionCases} used this module to calculate return-on-investment on energy arbitrage and frequency regulation services. The results showed that a higher number of arbitrage trades might lead to a lower cycle life of SLBs but generate a higher cash inflow and return on investment. The cost-benefit ratio generally decreases as the price of SLB drops.
Collectively, these studies outline the critical role of economic models, and the model should be selected based on the purpose of the study. However, it is challenging to compare the economic performance of various SLB applications across studies due to differences in the financial indicators and economic model selections. Therefore, standardization of economic evaluation methodologies for SLB-ESS is essential to showcase the advantages of using SLBs and promote their adoption in the market, while it is still absent in current studies \cite{Steckel2021ApplyingSystems}.

\subsubsection{Degradation-involved economic assessments for SLBs}

Before using retired batteries in the ESS, the remaining capacities of batteries need to be examined or estimated to initiate a safe and economical operation in second-life applications. As mentioned in Section \ref{sec:fastsc}, batteries with different state of health (SOH) levels would be available for second-life applications. Typically, SLBs with a higher remaining capacity yield more revenue, but they may come at a higher cost.
 
Exploring the degradation process of batteries helps optimize their usage and extend their lifespan. However, degradation quantification of SLBs is challenging because of the lack of reliable degradation models. Previous studies calculated the cost of degradation for both FLBs and SLBs by adding the replacement cost in economic analyses, assuming a fixed battery lifetime (e.g., five years), while ignoring the operating cost caused by battery degradation. However, a dynamic battery degradation process must be factored in when determining the profitability of SLBs since it harms the benefits gained by SLBs, as the benefits are generally proportional to the total amount of energy throughput across the battery lifespan, as indicated by the remaining useful life. Therefore, it is crucial to examine the battery degradation process under different operating conditions so that the revenue or cost savings brought by SLBs will at least cover their capital cost. Dynamic degradation models can help estimate the variable operating costs caused by the loss of energy due to battery capacity fade.
Recently, more and more studies have considered the dynamic degradation process when evaluating the economic performance of SLBs and recognized that battery aging has highly non-linear characteristics, with rapid capacity loss and resistance rise \cite{Song2019EconomyScenarios}. To incorporate the dynamic degradation process in the dispatching calculations or optimization process, researchers have employed simplification, linearization, and iteration technologies to fit existing cell-level battery degradation models. For example, previous studies have developed simplified semi-empirical linear degradation models using health factors \cite{Neubauer2011TheApplications}, constant degradation rates \cite{Steckel2021ApplyingSystems}, and lifetime throughput \cite{Debnath2016GridableGrid,Rallo2020lithium-ionCases,Neubauer2011TheApplications} to quantify battery degradation, while these simplified models are neither accurate enough for economic analysis nor able to reflect actual battery degradation mechanisms. Some studies linearized semi-empirical degradation models using techniques such as the Rainflow counting algorithm to approximate the cycling aging behavior of SLBs and incorporated them in their economic analysis \cite{Xu2018FactoringMarkets, Marques2019ComparativeFade}. Only a few studies have directly applied non-linear degradation models, such as the Arrhenius model, in their analysis as an offline degradation estimation method to get accurate estimations of capacity losses under specific mission profiles \cite{Song2019EconomyScenarios,Mathews2020TechnoeconomicAging, BAI2023233426}. There is a trade-off between the accuracy of battery degradation and the computational cost. Overall, studies that concluded that SLBs are profitable might have made overly optimistic hypotheses or assumptions regarding battery degradation behaviors in their second-life applications, leading to lower operational costs and longer service life.

\subsection{Other critical considerations}

\subsubsection{Cost estimation for SLBs}
Determining the pricing of SLBs is a critical factor in their market potential, while it is a challenging task. To mitigate the significant price uncertainties related to SLBs, three methods are used to determine their cost: the market survey, sensitivity analysis, and estimating the benefits of second-life use. Market surveys involve subtracting the refurbishment cost from the original price to determine the cost of SLBs \cite{Neubauer2012AValue,Neubauer2011TheApplications,Arshad2022LifeReview}. Some studies conducted a sensitivity analysis on the cost of SLBs and found the profitable purchase cost of SLBs \cite{Assuncao2016TechnicalEnergy,Fallah2021End-of-LifeModelling,Wu2020DoesBatteries,Bai2019EconomicChina}. Mathews et al. \cite{Mathews2020TechnoeconomicAging} assessed the profitability of a PV-SLBESS standalone system under different operating conditions and suggested that retailers sell SLBs for less than 60\% of the FLBs' price. Moreover, studies also estimated the benefits of FLBs during their SOH drops from 100\% to 80\% to determine the profit margin and appropriate cost of SLBs \cite{Song2019EconomyScenarios}. However, the cost considerations in existing studies are not precise, as the costs related to transportation, storage, and replacement are ignored, resulting in a lack of reliability in current techno-economic evaluations \cite{Sun2020EconomicChina,Steckel2021ApplyingSystems}. The uncertainties in repurposing costs, unreliable SLB market alignment, and lack of financial justification are the main barriers that impact decisions on deploying SLBs. Tighter cooperation between the industry and the research parties is the key to closing current information gaps, and the transparency of costs related to SLBs plays a significant role in improving the accuracy of economic assessments. The outcomes of such evaluations also provide critical information for industrial and governmental decision-makers regarding how SLBs should be implemented to achieve optimum benefits.

\subsubsection{Regional effects}
The profitability of SLBs is affected by various factors, including the price of electricity in different countries. For example, in China, industrial applications have a higher electricity price than the residential sector. Oppositely, in the US, the electricity price for residential applications is higher. This price difference can directly impact the profitability of SLBs, especially for applications like energy arbitrage. Gur et al. \cite{Gur2018TheAnalysis} mentioned that although PV-SLBESS was not profitable in European countries, the situation might be different in other countries.
External support, such as government subsidies, regulations, and market tariff policies, is critical for the economic feasibility of SLBs \cite{Huang2021EconomicChina,Gur2018TheAnalysis}. Policies and subsidies from governments and regulatory parties play a significant role in promoting SLBs \cite{Heymans2014EconomicLoad-levelling}, as verified by several existing studies, which reported that using SLBs can be profitable in power systems when governmental support and local tariffs are present \cite{Sun2020EconomicChina,MirzaeiOmrani2019EconomicSectors}. 

\subsubsection{Key assumptions}
In order to convert future cash flow into present value, a discount rate is applied, and its value mainly depends on the project risk, interest rate, and duration of the project. The value of the discount rate is mainly determined based on past studies, varying from 1\% to 15\%, as shown in Table 
The discount rate has a significant impact on the evaluation results \cite{Steckel2021ApplyingSystems,Kamath2020EconomicStorage}, as a higher discount rate reduces the future cash inflow \cite{Gur2018TheAnalysis}.
Another factor that plays a major role in assessing the benefit of SLBs is the EOL projection/assumption, i.e., the entire lifespan of batteries \cite{Zhu2021End-of-lifeBatteries}. Current studies assume a static SOH at EOL of around 50\%-60\% \cite{Song2019EconomyScenarios, Kamath2020EconomicStorage}, but this may not be accurate. It was found that the lower the SOH at EOL, the higher the benefit-cost ratio, since greater contributions from batteries accumulate over a longer period of time. So it is crucial to investigate the estimation of EOL points for SLBs in various second-life applications.

\section{Power Electronics Enabling Technologies}\label{sec:power}

\begin{figure}
    \centering
    \includegraphics[width = 10cm]{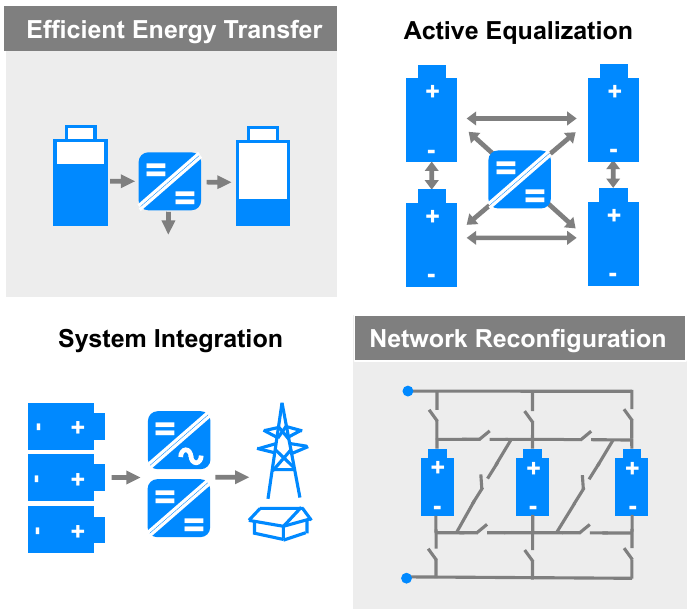}
    \caption{Core power electronic technologies that enable more efficient, reliable, and safer SL-BESS.}
    \label{fig:power_elec_overview}
\end{figure}

In a BESS, power electronics are the key energy conversion interfaces among the batteries, other distributed energy resources, loads, and power distribution networks.
The performance of BESS with SLBs (SL-BESS) is inadequate because the SLBs are low in power efficiency, heterogeneous in health conditions, and insufficient in reliability.
Well-designed power converters can substantially amend this performance inadequacy of SL-BESS:
(1) SLBs are heterogeneous in power capabilities, energy capacities, and health conditions.
Although the grouping can lower the heterogeneity, the costs of transportation and inventory in preparation for regrouping are significant.
Power converters are a beneficial alternative to the grouping because the converters can compensate for the discrepancy among SLBs instantaneously \cite{Abdel-Monem2017};
(2) SLBs are more unreliable and unsafe than new batteries. The power-electronics-based intelligent devices can swiftly protect the SL-BESS from detrimental damages such as catching fire and short circuits;
(3) SLBs and power converters are cascaded in an SL-BESS. Given the lossy SLBs, more efficient power conversion circuits and architectures can enhance the overall power efficiency of SL-BESS.
Thus, as illustrated in Figure \ref{fig:power_elec_overview}, we present a review of four power-electronics technologies that offer advantages to the SL-BESS: (1) efficient energy transformation; (2) active equalization; (3) network reconfiguration; and (4) system integration.

\subsection{Efficient energy transfer}
Different power electronics circuits are needed to transform the DC energy of batteries into DC, Pulse, and alternating current (AC) forms in the DC microgrid, pulse-charging of batteries, and AC grid, respectively. In all transformations, energy efficiency is a key performance indicator for battery storage systems.
SLBs are lossier due to the increase in resistance.
Moreover, SLBs are more complicated in the output impedance due to the SEI layer growth \cite{Roder2014}. This impedance significantly varies during the cycling process. 
By deploying more efficient power converters \cite{Schimpe2018}, the thermal management cost of the entire SL-BESS can be decreased, the energy efficiency improves, and battery life can be considerably extended.
Soft-switching techniques \cite{qian2011}, variable operation modes \cite{Mukherjee2016}, and design optimization \cite{Fang2015} have been utilized to improve the efficiency of power converters as well as the entire SL-BESS.

\subsection{Active equalization}
Equalization can improve the power and energy utilization of battery systems. The energy system with SLBs, which are heterogeneous in health conditions, can especially benefit from equalization.
\emph{Passive equalization} dissipates the excess energy as heat to homogenize the battery units.
The speed and efficiency of equalization can be significantly improved by the emerging \emph{active equalization} technologies.
In active equalization, the power flows among batteries are actively regulated by power converters.
Active equalization can uniform the energy, power, and SOH.
In active energy equalization, power electronics networks, together with state estimation and adaptive algorithms, can equally deplete the energy of all heterogeneous units at the same time \cite{Liu2020, Abdel-Monem2017}.
The ratings of the power converters can be customized so that the power capability of each battery unit is utilized and the power output of BESS can be maximized \cite{Cui2021a, Cui2022}.
The active SOH equalization leverages the information on the battery degradation dynamics.
With the information on the battery degradation dynamics, the power converters can drive the unhealthy battery unit less, drive the healthy unit more, and allow the degradation conditions of all units to converge \cite{azimi2021}.

In comparison to the existing literature that focuses mostly on cell-level equalization, this paper reviews the equalization architectures in a way that is agnostic to the battery cells, modules, and packs.
State-of-the-art equalization circuit architectures can be categorized into full power processing (FPP) and partial power processing (PPP).
\begin{figure}[htbp]
    \centering
    \includegraphics[width = 12cm]{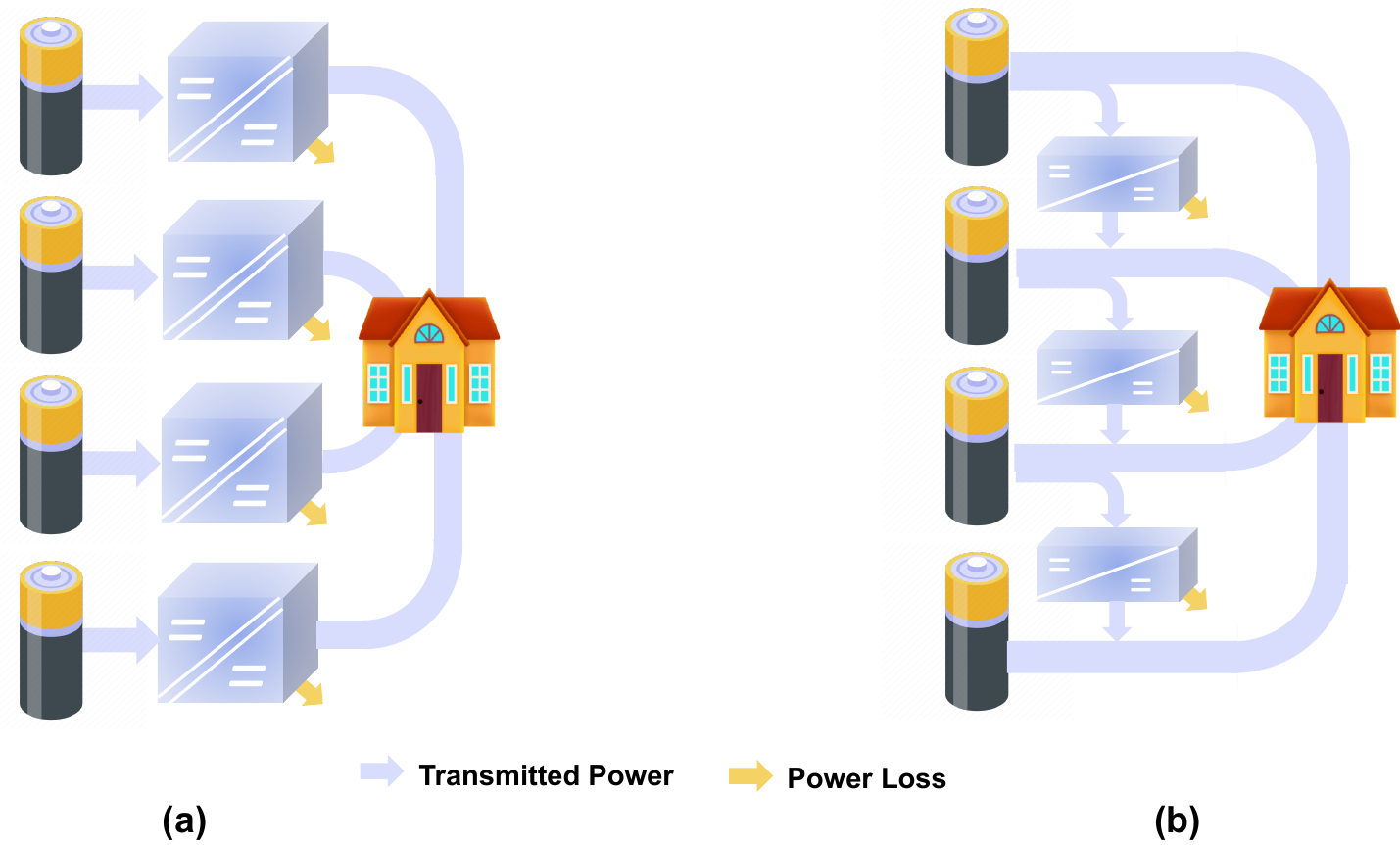}
    \caption{Different types of DC-DC power processing architectures: (a) partial power processing; (b) full power processing.}
    \label{fig:dcdc_partial_full}
\end{figure}
\begin{figure}[htbp]
    \centering
    \includegraphics[width = 12cm]{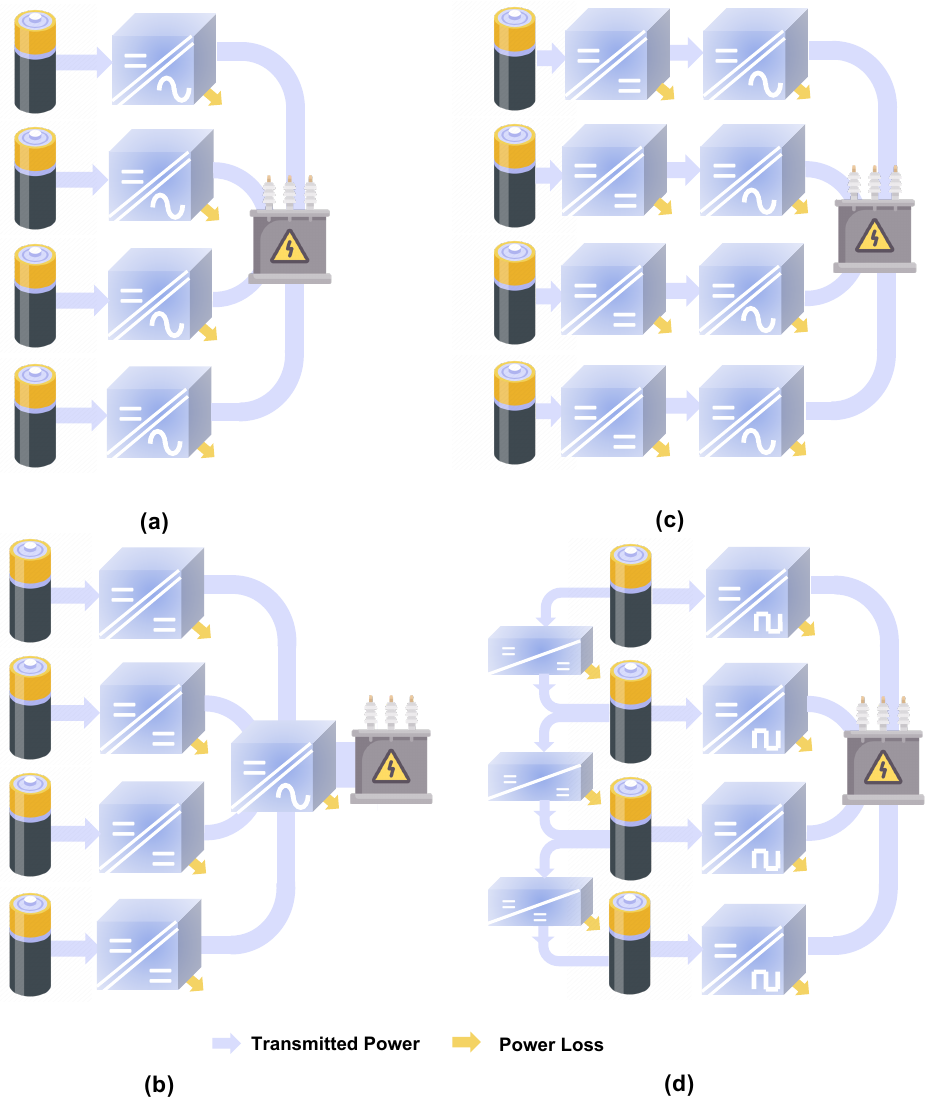}
    \caption{Different types of DC-AC power processing architectures: (a) one-stage full power processing; (b) two-stage full power processing with a bulk inverter; (c) two-stage full power processing with microinverters; (d) one-stage partial power processing.}
    \label{fig:dc_ac}
\end{figure}
\subsubsection{Full power processing}
In FPP, each battery unit delivers power to the load through a power converter.
The power that permeates the SLBs needs to flow through the power converters.
The power converter is designed to compensate for the heterogeneous electrical characteristics of SLBs.
The circuit architectures of FPP will be reviewed in both DC-DC and DC-AC conversions as shown in Figures \ref{fig:dcdc_partial_full} and \ref{fig:dc_ac}, respectively.
In both conversions, the SLB-converter units can be connected in series or parallel \cite{Maughan2022} to the output.

In the DC-DC conversion, each SLB unit is connected to the DC bus through a DC-DC converter, as shown in Figure \ref{fig:dcdc_partial_full}(a).
The topology of DC-DC converters varies with the performance specifications \cite{Mukherjee2015a, Maughan2022}. Mukherjee et al.  \cite{Mukherjee2015a} used the boost topology because it can step up the low battery voltage, decrease the number of series batteries, and carry a small ripple current to batteries.
The work by \cite{Maughan2022} discussed an isolated $LC$-resonant topology for wider output step-up range, higher power efficiency, and galvanic isolation.
Two architectures are typically used in DC-AC conversion. One architecture is a two-stage that includes a DC-DC stage and a DC-AC stage, as shown in Figure \ref{fig:dc_ac}(b).
The first DC-DC stage is the same as the DC-DC conversion and the second DC-AC stage usually contains a bulk inverter \cite{Mukherjee2015a}.
The other architecture is the series or parallel combinations of microinverters \cite{Mukherjee2014, Ma2020a}, as shown in Figure \,\ref{fig:dc_ac}(a).
A DC-DC converter can be cascaded in the front of each modular inverter so that the power flow of each battery can be individually regulated \cite{Mukherjee2012, Mukherjee2014, Ma2020a}, as shown in Figure \ref{fig:dc_ac}(c).

\subsubsection{Partial power processing}
In PPP, the major power flow is directly delivered from batteries to the load. The power converters only process the power mismatch among batteries, therefore the energy utilization and efficiency of the system are increased. 
Most PPP reported in the literature is for series-output, DC-DC conversion architecture \cite{Milas2022, Wang2018c, Mubenga2021}. According to the layers of power conversion networks, PPP can be classified as flat PPP and hierarchical PPP. 
Flat PPP architecture contains a single layer of power conversion network, as shown in Figure \ref{fig:dcdc_partial_full}(b).
There are a variety of implementations of the flat PPP architecture for DC-DC conversion \cite{Milas2022, Wang2018c, Mubenga2021, Milas2022, Wang2018c, Mubenga2021, Ziegler2019, Hua2020, Mukherjee2014, Ma2020a}.
The research in \cite{Milas2022, Wang2018c, Mubenga2021} connected a power converter to two adjacent batteries to exchange the mismatched energy. 
In \cite{Einhorn2011a, Imtiaz2013a, Einhorn2011b, Nguyen2022}, the authors connected a converter between each battery and the output bus. Compared to the FPP, this converter only processes the power mismatch between the battery and the output bus.
\cite{Ziegler2019} proposed a virtual bus architecture that substitutes this output bus with an extra DC bus that is formed by a capacitor. 
In \cite{Hua2020}, batteries exchange energy through an AC bus.
A few papers have investigated the PPP solutions for parallel SLB units. For example, La et al. \cite{La2020} used the dynamical resistor to control the current of each SLB unit.
The papers that demonstrated PPP for DC-AC conversion majorly utilized Multilevel Modular Converters (MMC) architectures, as shown in Figure \ref{fig:dc_ac}(d), and adaptive battery balancing algorithms \cite{Mukherjee2014, Ma2020a}.

The performance of flat PPP degrades as the heterogeneity of SLBs grows. This motivates the hierarchical PPP architectures, where multiple layers of power converters are utilized to equalize the heterogeneity among not only neighbor batteries but also far-end batteries \cite{Zhang2017a, Cui2022, Cui2021a, Cui2021c}.
Zhang et al. \cite{Zhang2017a} demonstrated that a two-layer series-output hierarchical PPP can reduce energy loss, lower the current rating, and decrease the cost and volume of the system.
The design of hierarchical PPP is complicated. 
The work in \cite{Cui2022} and \cite{Cui2021a} proposed a systematic data-driven design methodology that is based on the statistical distribution of SLB supply. The power converter cost and system power efficiency are significantly decreased compared to the flat PPP.
Cui et al. \cite{Cui2021c} derived a parallel-output hierarchical PPP for managing the SLB packs as EV charging buffering.

\subsection{Network reconfiguration}
SLBs are more dispersed in their capacity and capability. The life of the whole SL-BESS should not be affected by the earlier termination of a unit. Converters should be designed so that the weakest reliability link is not on power converters.
In the existing literature, redundancy, dynamic reconfigurability, and hot-swapping technologies have been used to enhance the reliability of SL-BESS. Mukherjee et al. \cite{Mukherjee2014} created redundancy by using a ``k-out-of-n'' system, where the SL-BESS can operate as long as k battery-power converter units out of n modules are working properly. The work by \cite{Ci2016} proposed an adaptive reconfigurable multi-cell structure to improve reliability. Maughan et al. \cite{Maughan2022} proposed a hot-swapping approach that allows the whole system to operate seamlessly and without any interruption when battery packs need to be disconnected in faults.

SLBs are highly degraded and more vulnerable to thermal runaway, short circuits, and overcharge.
The power-electronics-based intelligent devices play major roles in fault protection, diagnostics, and prognostics.
From the fault protection perspective, as exhibited in \cite{Ci2016}, the solid-state breakers can break the short-circuit faults much faster than the mechanical contractors.
From the prognostics perspective, as demonstrated in \cite{Liu2021, Oji2021}, the power conversion network can be dual-used to perform the online Electrochemical Impedance Spectroscopy and online SOH so that the degradation condition is monitored.

\subsection{System integration}
The major application of SL-BESS is in stationary energy storage systems \cite{song_benefit_2022}.
According to the mission profiles that the SL-BESS performs in the system, SL-BESS can be classified into four categories: energy-intensive DC-DC conversion, energy-intensive DC-AC conversion, power-intensive DC-DC conversion, and power-intensive DC-AC conversion.

In energy-intensive DC-DC conversion, SL-BESS colocated with the solar can support DC microgrids \cite{Hamidi2013, Deng2021, Fares2019}. The energy-intensive DC-AC conversion case, for example, includes the SLBs directly integrated into the AC power grid or AC microgrids for the services of peak shaving, spinning reserve, or renewable firming \cite{Hossain2019, Silvestri2021, Heymans2014}. In both cases, the system-level control question is whether SLBs with different energy levels can be proportionally utilized and depleted simultaneously.

In power-intensive DC-DC conversion, SLBs act as a buffer for the EV's fast charging \cite{DArpino2019, Hamidi2013, Deng2021}. The power-intensive DC-AC conversion case, for example, includes the SLBs integrated with the grid-forming inverters for transmission stabilization \cite{Hossain2019, Silvestri2021} as well as frequency regulation \cite{Janota2020} and uninterrupted power supply for public building backups \cite{Milas2022}. In both cases, the system-level control question is whether BESS can push high enough power support in a short period of time.

\section{Conclusions}\label{sec:con}
Due to the high volume of EVs being in service and the limited lifetime of the onboarding LIBs, the large-scale retirement of LIBs will come in the near future. Since SLBs retired from EVs still own 70\%-80\% of the initial capacity, they have the potential to be utilized in scenarios with lower energy and power requirements. In this way, the value of LIBs can be maximized in their second-life applications. 
This paper reviews critical technologies and the economics of SLBs. 

The degradation models for FLBs mainly include PMBs, DDMs, and hybrid models. While physics-based models are explainable, they cannot be applied to situations where the degradation modes are not well understood. In contrast, DDMs circumvent this issue by leveraging high-quality data, but their interpretability is limited. Hence, there is an opportunity to apply hybrid models to establish explainable models with physical laws that can be generalized. The missing historical data over the first-life operation and poorly understood knee points have posed two challenges to accurately modeling the degradation behaviors of SLBs. At present, there is a dearth of degradation models specifically designed for SLBs. Therefore, we investigate the feasibility of utilizing battery degradation models developed for FLBs in the context of SLBs.
In addition, the retired batteries disassembled from the battery pack will experience fast screening and regrouping to be applied to second-life applications. The fast screening includes visual inspection and fast parameter checks. Given the anticipated high volume of used batteries in the future, there is a need to develop an automated machine that can efficiently disassemble and screen batteries using AI techniques. Additionally, this paper presents, for the first time, a regrouping standard that can guide the classification procedure and enhance the performance and safety of SLBs.
For economic considerations, only those profitable SLB applications are worth further promotion. Therefore, it is crucial to conduct a thorough economic analysis of SLBs prior to their implementation in second-life applications.
This paper provides a summary of the profitable applications of SLBs and those that hold significant potential. Specifically, SLBs have proven to be beneficial in a variety of applications, including renewable firming, area and frequency regulation, and peak shaving. Additionally, the integration of SLBs in micro-grid and smart-grid applications has also shown feasibility.
 
Finally, power electronics serve as the critical energy conversion interfaces among batteries in SLB applications, and appropriately designed power converters have the potential to significantly enhance the performance of SLB applications in terms of power, electrical characteristics, and reliability. This paper provides a comprehensive review of advanced power electronics technologies, including high-efficiency energy transformation, active equalization, network reconfiguration, and system integration.

\section*{Declaration of Competing Interest}
The authors have no conflict of interests to declare about this work.

\section*{Acknowledgment}
This work is supported by the National University of Singapore Start-Up Grant (A-0009527-01-00).




\bibliographystyle{elsarticle-num} 
\bibliography{all_bib}


\end{document}